\documentclass[twocolumn,showpacs,prb,amsfonts,amsmath,amssymb,floatfix]{revtex4}

\usepackage{hhline}
\usepackage{mathrsfs}
\usepackage{graphicx}
\usepackage{dcolumn}
\usepackage{bm}

\input{epsf}

\begin{document}

\title{
Spin density functional study on magnetism of potassium loaded Zeolite A
}

\author{Yoshiro Nohara$^{1}$}
\author{Kazuma Nakamura$^{2}$}
\author{Ryotaro Arita$^{2}$}
\affiliation{$^1$Department of Physics, University of Tokyo, Tokyo 113-0022, Japan} 
\affiliation{$^2$Department of Applied Physics, University of Tokyo, Tokyo 113-8656, Japan} 

\date{\today}

\begin{abstract}
In order to clarify the mechanism of spin polarization in
potassium-loaded zeolite A, we perform {\em ab initio} density-functional
calculations. We find that (i) the system comprising only non-magnetic
elements (Al, Si, O and K) can indeed exhibit ferromagnetism, (ii) while
the host cage makes a confining quantum-well potential in which $s$- and
$p$-like states are formed, the potassium-4$s$ electrons accommodated in the
$p$-states are responsible for the spin polarization, and (iii) the size
of the magnetic moment sensitively depends on the atomic configuration of
the potassium atoms. We show that the spin polarization can be described
systematically in terms of the confining potential and the crystal field
splitting of the $p$-states.

\end{abstract} 
\pacs{73.22.-f, 75.75.+a, 82.75.Vx}
\keywords{Zeolite}
\maketitle 

Searching for novel ferromagnets comprising only non-magnetic elements 
has been a fascinating challenge in condensed matter physics. Up to present,
motivated by fundamental interests or potential technological importance, 
various ferromagnets with non-magnetic elements have been 
synthesized.\cite{Allemand,Takahashi} Among them, the alkali-metal-loaded zeolite 
is certainly a unique ferromagnet. Depending on the crystal structure of the host 
cage and the number/species of the guest cluster atoms, it exhibits not only 
ferromagnetism but also a rich variety of 
magnetic properties.\cite{Recent_Nozue,KLTA} In fact, one can envisage to design and control the magnetic 
properties of this system by choosing appropriate combinations of the guests 
and hosts.\cite{Arita}

When we consider such materials design, the first step we should take is to clarify
the mechanism of the spin polarization in this system. Indeed, it is of great interest 
to consider why clusters of alkali atoms confined in zeolite cages can be magnetic,
even though the bulk alkali metals are usually non-magnetic. However,
while almost 20 years have passed since the seminal discovery of the spin-polarized 
ground state in zeolite A,\cite{KLTA} 
{\em ab initio} studies on magnetism in zeolitic materials have been quite limited so far. 
This is mainly because (i) the electronic structure is expected to be very complicated 
since the unitcell of the system is extremely huge [typically it has $O(100)$ atoms and 
the lattice constant is $\sim$ 30 \AA], and (ii) the atomic configuration has not been 
determined accurately in experiments due to the extreme complexity in the structure 
of the system. 

However, fortunately, the situation of potassium-loaded zeolite A 
(K$_{12+n}$Si$_{12}$Al$_{12}$O$_{48}$, hereafter we call it as K-LTA) is different; 
a detailed neutron powder diffraction study has been performed \cite{Neutron} and thus we exceptionally 
have a reliable reference for {\em ab initio} calculations. On top of that, recent density 
functional calculations based on local density approximation (LDA)
have clarified that the low-energy electronic structure of this compound is quite simple.\cite{Arita}
Namely, the system is regarded as a {\em supercrystal}, where guest potassium 
clusters act as a {\em superatom} with well-defined $s$- and $p$-like orbitals.\cite{Cluster} 
Thus, we may expect that the low energy physics is expected to be described 
systematically in terms of a superatom.

Experimentally, the spin polarization in K-LTA is observed when we introduce more than two 
potassium atoms (i.e., $n>2$) and the Curie and Weiss temperature take their maximum 
at $n\sim 4$. If we follow the picture mentioned above, the superatom $s$ state 
is fully occupied and three $p$ states are partially filled for $n\sim4$.  
While several scenarios for the spin polarization in K-LTA based on the assumption 
that the orbital degeneracy of the $p$ states plays a crucial role have been 
proposed,\cite{SpinCant,Maniwa,Nakano} an {\em ab initio} calculation based on 
spin density functional theory has yet to be done.
The purpose of the present study is to clarify the mechanism of spin polarization 
in K-LTA from first principles. We examine in detail (i) whether a ferromagnetic ground state 
is really realized just from Si, Al, O and K, and (ii) if it does, for which condition the
system has spin polarization. Hereafter, we focus on the case of $n=4$, where the energy scale 
of magnetism is expected to be highest. 

\begin{figure}[htbp]
  \vspace{0cm}
  \begin{center}
  \includegraphics[width=0.45\textwidth]{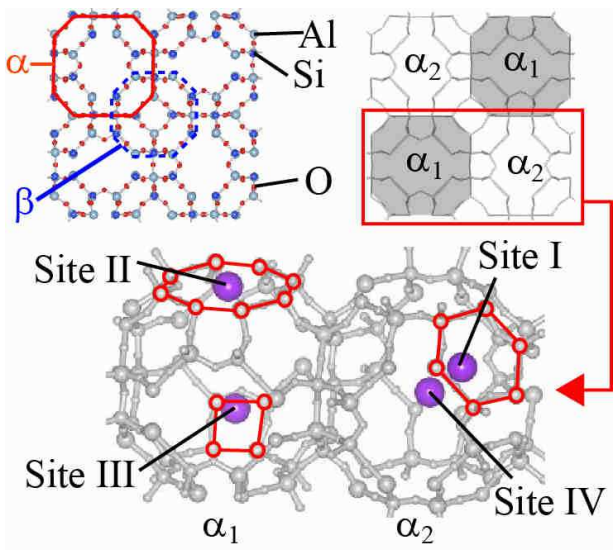}
  \caption{(Color online) 
Upper left: Overall profile of zeolite LTA, where O, Al, Si atoms are 
denoted by red-small, light-blue, and dark-blue spheres, respectively. 
$\alpha$ and $\beta$ cages are marked by red-solid and blue-dashed lines, 
respectively.
Upper right: Two kinds of $\alpha$ cages ($\alpha_1$ and $\alpha_2$). 
Lower panel: Four kinds of sites occupied by potassiums (purple-large spheres).
For the definition of the sites I-IV, see the text.}
  \label{fig_LTA}
  \end{center}
\end{figure}

Let us see the detail of the atomic configuration considered in the present study. 
Figure \ref{fig_LTA} shows the atomic geometry determined by the neutron measurement with
an assumption that the system has the symmetry of space group F23.\cite{Neutron} 
Due to the Lowenstein rule (Si and Al should be arranged alternately), the unitcell 
contains two large (small) alminosilicate cages which we call $\alpha$ ($\beta$) cages 
(upper left panel). While a simplified unitcell having one $\alpha$ and $\beta$ cage 
(which violates the Lowenstein rule) is employed in the previous LDA calculation,\cite{Arita} 
here we adopt the original unitcell having $\sim$200 atoms.
Hereafter, we label the two $\alpha$ cages as $\alpha_1$ and $\alpha_2$ (upper right panel). 
While the positions of Al, Si, and O were uniquely determined by the measurement, 
the experiment found several possible positions for potassium atoms in the $\alpha$ cages (bottom panel).
Following this observation, we denote the center of six-, four-, and eight-membered rings 
in the $\alpha$ cage as site I, II, and III, respectively. The site IV is the center of the $\alpha$ 
cage. In order to indicate which $\alpha$ cage the site I, III, and IV belong to, we label them 
with subscript such as site I$_{\alpha_1}$ or I$_{\alpha_2}$. Notice that the site II is on 
the border of two $\alpha$ cages.

The experimental occupation numbers of site I$_{\alpha_1}$, I$_{\alpha_2}$, 
II, III$_{\alpha_1}$, III$_{\alpha_2}$, IV$_{\alpha_1}$, and IV$_{\alpha_2}$ 
are 8.0, 8.0, 6.4, 5.9, 3.5, 0.0, and 0.5, respectively. In the present 
calculation, for simplicity, we assume the occupation numbers as 8, 8, 6, 6, 3, 0, 
and 1. Thus the site I, II and IV$_{\alpha_2}$ are fully occupied and the 
site IV$_{\alpha_1}$ is completely vacant.\cite{site-I}
Note that the total number of the potassium atoms is 32, and the number of 
site I, II, III in the unitcell is 16, 6 and 24, respectively. 
While the atomic configuration of the site I, II and IV are uniquely determined, 
there are huge possibilities for the configuration of the site III;
since there are 24 possible positions and we have nine potassium atoms in the 
site III (6 for the $\alpha_1$ cage and 3 for the $\alpha_2$ cage),
we have $_{12}C_{6}$$\times$$_{12}C_{3}$=203280 possibilities.

\begin{figure}[htbp]
\vspace{0cm}
\begin{center} 
\includegraphics[width=0.45\textwidth]{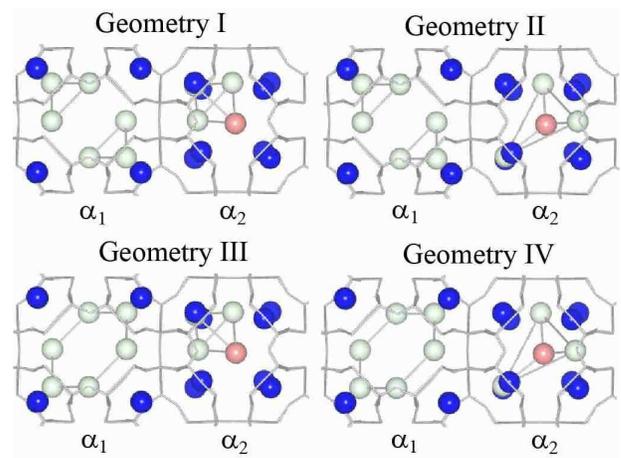}
\caption{(Color online) 
Four geometries considered in the present study, where dark-blue, light-green, 
and medium-red spheres stand for the potassiums occupying the site I, III, 
and IV, respectively. The difference in each geometry comes from the difference 
in the configuration of site-III potassiums; 
 ``two triangles" and ``small triangle" (geometry I), 
``two triangles" and ``large triangle" (geometry II), 
``hexagon" and ``small triangle" (geometry III), 
and ``hexagon" and ``large triangle" (geometry IV) are accommodated 
in the $\alpha_1$ and $\alpha_2$ cages. We note that four site-I potassiums 
in the $\alpha_1$ cage are overlapped with the remaining four site-I 
potassiums in this view.}
\label{fig_FourType}
\end{center}
\end{figure}

Thus in the present study, we focus on the following four geometries I-IV with
trigonal symmetry shown in Fig.~\ref{fig_FourType}.
In those geometries, we have two triangles and a hexagon in
the $\alpha_1$ cage, and a small triangle and a large triangle
in the $\alpha_2$ cage. We note that, for the geometries I and III, 
the site-III and site-IV potassiums in the $\alpha_2$ cage make a tetrahedron. 

{\em Ab initio} calculations based on local spin density approximation (LSDA) 
were performed with {\em Tokyo Ab initio Program Package}.\cite{TAPP}
We used the GGA exchange-correlation functional,\cite{GGA} plane-wave basis set, 
and the ultrasoft pseudopotentials \cite{US} in the Kleinman-Bylander representation.\cite{PP} 
The pseudopotential of potassium was supplemented by the partial core correction.\cite{PCC} 
The energy cutoff of plane waves representing wavefunctions is 36 Ry and a 
4$\times$4$\times$4 $k$-point sampling is employed.

We show in TABLE~\ref{tab_FourType} the total energies calculated for the geometries I-IV, 
together with local magnetic moments in each $\alpha$ cage.\cite{mu} 
We see that the site-III configuration significantly affects the energetics and the value 
of the magnetic moment. 
Interestingly, for all the geometries, the moment develops in the $\alpha_2$ cage, 
which suggests that the valence electrons responsible for the magnetism reside mainly 
in the $\alpha_2$ cage. This is because the confining potential of the $\alpha_2$ cage 
is deeper than that of $\alpha_1$; as seen from Fig.~\ref{fig_FourType}, the site-I potassiums 
in the $\alpha_2$ cage exist at inner side than those in the $\alpha_1$ cage, 
which makes the electrostatic potential in the $\alpha_2$ cage deeper.
\begin{table}[bhtp]
\caption{Total energy per unitcell of each geometry $\Delta E_{{\rm tot}}$ with reference 
to that of the geometry I and the local magnetic moments $M_i$ in the $\alpha_i$ 
cage ($i$ = 1 or 2) (Ref.~\onlinecite{mu}). For the definition for the 
geometries I-IV, see Fig.~\ref{fig_FourType}.}
\label{tab_FourType}
\begin{center}
\begin{tabular}{r@{\ \ \ \ }r@{\ \ \ \ }r@{\ \ \ \ }r@{\ \
 \ \ }r}
\hline \hline
  & Geom I & Geom II & Geom III & Geom IV \\ \hline
$\Delta E_{{\rm tot}}$(eV)&  0  &    1.02   & 1.49  &    0.12 \\
$M_{1}$($\mu_{B}$) & $-$0.02    & $-$0.20   & 0.06  & $-$0.16 \\
$M_{2}$($\mu_{B}$) &    1.92    &    0.33   & 1.36  &    1.00 \\ \hline \hline
\end{tabular}
\end{center}
\end{table}
\begin{figure}[htbp]
  \vspace{0cm}
  \begin{center}
  \includegraphics[width=0.45\textwidth]{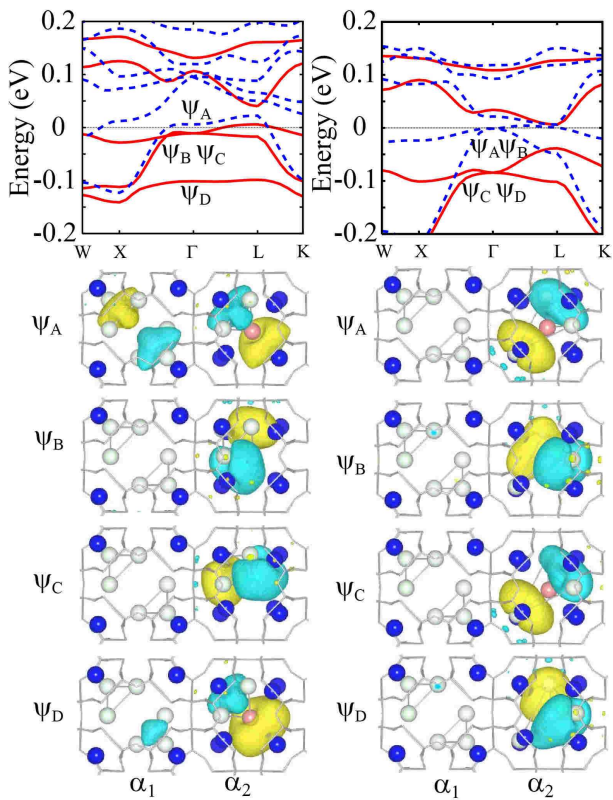}
  \caption{(Color online) 
Calculated GGA band dispersions (upper) and Bloch wavefunctions, 
$\Psi_{\rm A}$-$\Psi_{\rm D}$, at the $\Gamma$ point (lower) for 
the geometries I (left) and II (right). In the band dispersion, 
red-solid and blue-dashed lines stand for majority and 
minority spin states, respectively. The energy zero is set to the 
Fermi level. In the wavefunctions, isosurfaces are drawn by the 
values of $\pm$0.015 (a.u.). Potassiums occupying the site I, III, 
and IV are displayed by spheres of the same colors with 
Fig. \ref{fig_FourType}. 
}
  \label{fig_band}
  \end{center}
\end{figure}

Another important point in TABLE~\ref{tab_FourType}
is that the size of the magnetic moment depends on the atomic configuration
of the potassium cluster. In order to clarify its origin, we calculated 
the low-energy bands and their wave functions. In the upper left (right) panel 
in Fig.~\ref{fig_band}, we show the low-energy band structures near the Fermi 
level for the geometries I (II), corresponding to the second (third) column 
in TABLE~\ref{tab_FourType}.
While the band structure is quite different for these two geometries, 
the Bloch wavefunctions at the $\Gamma$ point ($\Psi_{\rm A}$-$\Psi_{\rm D}$) have 
common features; (i) they have large amplitude in the $\alpha_2$ cage due to the deeper 
cage potential in $\alpha_2$ mentioned above, and (ii) they widely spread in the $\alpha$ 
cage without localizing on any specific atoms and are regarded as {\em superatom} 
$p$-type wavefunctions.\cite{Arita}

\begin{figure}[b]
  \vspace{0cm}
  \begin{center}
  \includegraphics[width=0.45\textwidth]{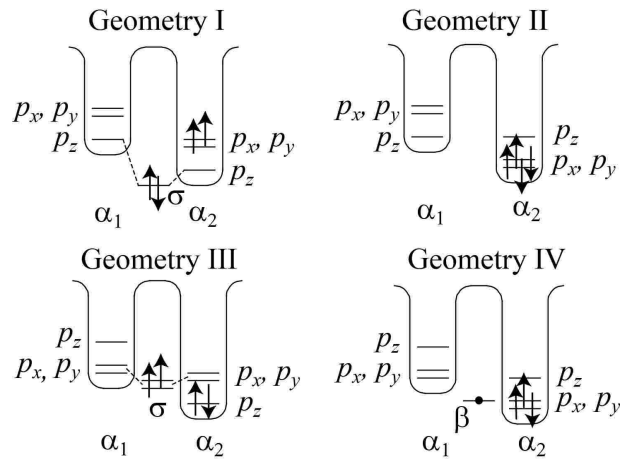}
  \caption{Level diagrams for electronic structures of the four geometries displayed in 
Fig.~\ref{fig_FourType}, where superatom $p$ levels in the $\alpha_1$- and $\alpha_2$-cage 
potentials and the resulting spin configurations are drawn schematically. 
The $p$ states are represented in the local coordinates (for the definition, see the text). 
The difference between the cage potentials of $\alpha_1$ and $\alpha_2$
comes from the difference in the configuration of the site-I potassiums 
(see Fig.~\ref{fig_FourType}) and the $p$-level splitting is due to the 
site-III-potassium configurations.}
  \label{III_IV}
  \end{center}
\end{figure}

To make the situation clearer, in Fig.~\ref{III_IV}, we draw level diagrams of the 
low-energy states where the cage potentials, superatom $p$ levels, and spin 
configurations are depicted schematically. 
Here, the $p$ states are represented in the local coordinates, where 
the $z$ axis is taken to be perpendicular to the triangles in $\alpha$ cages (see Fig.~\ref{fig_FourType}). 
In the geometry I (upper left), for both the $\alpha$ cages, the superatom $p_x$ and $p_y$ 
levels are degenerated and higher than the $p_z$ level. 
This can be understood as follows:
In the $\alpha_1$ cage, the superatom $p_z$ lobe extends in the direction
of the center of the triangle and feels a positive crystal field, lowering the level of $p_z$. 
The same mechanisum works in the $\alpha_2$ cage, and the low-lying $p_z$ states in 
$\alpha_1$ and $\alpha_2$ cages form a $\sigma$ bonding orbital. On the other hand, 
the $p_x$ and $p_y$ states in the $\alpha_2$ cage, with a non-bonding character, reside near 
the Fermi level. In the present case of $n$ = 4, there are four superatom 
$p$ electrons. Two electrons accommodated in the $\sigma$ bonding orbital 
and the remaining two electrons occupy the degenerated $p_x$ and 
$p_y$ orbitals.  According to the Hund's rule, electron spins in the 
degenerated orbitals align to be parallel, thus generate the magnetic 
moment of $\sim$ 2 $\mu_{{\rm B}}$. 

For the geometry II (upper right panel in Fig.~\ref{III_IV}), in the $\alpha_2$ cage, 
the level of $p_z$ is higher than $p_x$ and $p_y$. This can be also understood in terms 
of the crystal field formed by the site-III potassiums in the $\alpha_2$ cage. 
In this configuration,
the $p_x$ and $p_y$ orbitals lie in the ``large" triangle plane and thus these 
states are stabilized electrostatically, compared to the $p_z$ state. 
Consequently, in the geometry II, the four electrons fully occupy 
the $p_x$ and $p_y$ states, so that the system has smaller magnetic moments.

A similar argument can hold also in the cases of 
the geometries III and IV. In these geometries, the $\alpha_1$ 
cage contains a hexagon instead of two triangles.
The crystal field in the $\alpha_1$ cage stabilizes the $p_x$ and $p_y$ states 
in the hexagon plane. In the geometry III (lower left panel in 
Fig.~\ref{III_IV}), the two electrons occupy the low-lying $p_z$ orbital 
in the $\alpha_2$ cage and the degenerated $\sigma$ orbitals 
made from the $p_x$ and $p_y$ states in the $\alpha_1$ and $\alpha_2$ cages 
accommodate the remaining two electrons. Because the $\sigma$ orbital 
is less localized, the value of the moment is somewhat reduced 
($\sim$ 1.5 $\mu_{{\rm B}}$), compared to that of the geometry I.  
In the geometry IV (lower left panel), an extra state in
the $\beta$ cage appears near the Fermi level. 
This state 
has a large dispersion
and small spin polarization. The net moment is generated 
from the three electrons occupying the degenerated $p_x$ and $p_y$ orbitals, 
with the value of the moment $\sim$ 1 $\mu_{{\rm B}}$.

Thus, we conclude that K-LTA can be ferromagnetic in LSDA and the mechanism 
of the spin polarization 
is consistently explained in view of the superatom picture; if the $p$ levels
formed by the superatom wavefunctions are degenerated and partially occupied,
the system becomes magnetic, due to the Hund's rule coupling in these states.
This condition and the size of the moment are controlled by the atomic 
configuration of the guest potassium clusters, responsible for the cage-potential depth
and the crystal field inducing the $p$-level shift and split, respectively.

Experimentally, two kinds of possible magnetic structures have been proposed for K-LTA. 
One is the ferrimagnetic model,\cite{Maniwa} and the other is the spin-canted 
antiferromagnetic model.\cite{SpinCant}
In the former, the system consists of two sublattices with
large ($\sim$ 2.8 $\mu_{{\rm B}}$) and small (almost 0.0 $\mu_{{\rm B}}$)
magnetic moment, while, in the latter, every $\alpha$ cage has a moment of 1 $\mu_B$. 
These magnetic structures look like the results for the geometries I and IV, 
although the size of the moments is relatively smaller than the experimental values. 
It is interesting to note that the geometries I and IV 
 have comparable energies (see TABLE~\ref{tab_FourType}). 

In reality, the site III is randomly occupied.\cite{Neutron}
It is formidable or almost impossible to take into account the randomness of the site III 
explicitly in the first-principles calculation.
However, the superatom 
picture has been shown to capture essential features of the low energy physics 
of K-LTA.
Thus, a multi-orbital Hubbard-type model with random potentials 
would be served as a realistic low-energy model describing magnetism of the system.
Since all the parameters characterizing the Hubbard model can be derived with 
first-principle basis as was recently done in Ref. \onlinecite{Nakamura},
we can discuss the magnetism in more realistic situations by solving this effective 
model. It will be an important future study.

To summarize,
by means of {\em ab initio} spin density functional calculation, 
we found that (i) K-LTA comprising only Al, Si, O and K can be ferromagnetic, 
(ii) the superatom $p$ states are responsible for the magnetism,
and (iii) the magnetic property of K-LTA depends sensitively on the atomic configuration 
of the guest potassium cluster. The condition for the spin polarization was 
presented and understood systematically in terms of the superatom picture. 
Our findings provide a firm basis and will be the first step for clarifying 
the low-energy physics in zeolitic materials.

We thank Professor Yasuo Nozue, Takehito Nakano, and Mutsuo Igarashi 
for fruitful discussions. This work was supported by Scientific Research 
on Priority Areas of New Materials Science Using Regulated Nano Spaces (No. 19051016) MEXT, 
Japan. All the computations have been performed on Hitachi SR11000 system at 
the Supercomputing Division, Information Technology Center, the University of Tokyo.


\begin{thebibliography}{99}
\bibitem{site-I}
In the experiment, the exact position of a half of the site-I potassiums is slightly different from the center of the six-membered ring and two positions for the site I have been observed. The occupancies are different (57 \% and 43 \%). In the present study, we employ the high-probability position for all the site-I potassiums.
\bibitem{mu}
The local magnetic moment of $\alpha$ cages was calculated for the Voronoi polyhedron having the same volume as that of the sphere of the radius of 7.64 \AA.
\end{thebibliography}
\end{document}